\documentclass[letterpaper, 10 pt, conference]{ieeeconf}  

\IEEEoverridecommandlockouts                              
\usepackage{graphics} 
\usepackage{xcolor}
\usepackage{epsfig} 
\usepackage{amsmath} 
\usepackage{amssymb}  
\usepackage{cite}
\usepackage{soul}

\title{\LARGE \bf
Slow Inter-area Electro-mechanical Oscillations Revisited:\\ Structural  Property of Complex Multi-area Electric Power Systems}

\author{\thanks{This material was supported by the {MIT Aero-Astro Distinguished Scholars Program}  and  the NSF Grant No. ECCS-2002570 entitled ASCENT: Boosting Cyber and Physical Resilience of Power Electronics-Dominated Distribution Grids.The authors also greatly appreciate discussions with Dr. Rupamathi Jaddivada.} 
\thanks{$^1$ Hiya Akhil Gada hiyagada@mit.edu is a Ph.D. candidate at the Laboratory for Information and Decision Systems (LIDS) of Massachusetts Institute of Technology, Cambridge, MA, USA. {\tt hiyagada@mit.edu}}
\thanks{$^2$ Marija D. Ilic is a Joint EECS Adjunct Professor and a Senior Research Scientist at Laboratory for Information and Decision Systems (LIDS), Massachusetts Institute of Technology (MIT), Cambridge, USA. She is also  a Professor Emerita in the Electrical and Computer Engineering (ECE) Department of Carnegie Mellon University, Pittsburgh, PA. {\tt ilic@mit.edu}}  Hiya Akhil Gada$^1$ and  Marija D. Ilic$^2$}

\begin{document}

\maketitle
\thispagestyle{empty}
\pagestyle{empty}

\begin{abstract}
This paper introduces a physically-intuitive notion of inter-area dynamics  in systems comprising multiple interconnected energy conversion modules. 
The idea builds on an earlier general approach of setting their structural properties  by modeling internal dynamics in stand-alone modules (components, areas) using the fundamental conservation laws between energy stored and generated, and then constraining explicitly their Tellegen's quantities (power and rate of change of power). In this paper we derive, by following the same principles,  a  transformed  state-space model  for a general nonlinear system.  
Using this model we show the existence of an area-level  interaction variable, $intVar$,  whose rate of change depends solely on the  area internal power imbalance {and is independent of the model complexity used for representing individual module dynamics in the area}.   
Given these structural properties of stand-alone modules, we define in this paper for the first time an inter-area variable  as the difference of  power wave incident to tie-line from Area I and the power reflected into tie-lie from Area II.  
 Notably, these power waves represent the {interaction variables} associated with the two respective interconnected  areas.
We illustrate these notions using  a linearized case of two lossless inter-connected areas, and  show the existence of a new  inter-area mode when the areas get connected. We suggest that lessons learned in  this paper  open possibilities for computationally-efficient modeling and control of inter-area oscillations, and offer further the basis for modeling and control of dynamics in changing systems comprising faster energy conversion processes. 

\end{abstract}

\section{Introduction}


The existence of low-frequency electro-mechanical oscillations  has been recognized and studied as  a long-standing industry problem \cite{Kundur, Vijay},
which has recently become even more pronounced. To mention a few, electric power systems have experienced these problems in the form of  sub-synchronous resonance (SSR) between turbine modules and series compensated transmission lines  \cite{break}; low-frequency oscillations between different control areas during  loss of large power plants \cite{king}; and, more recently, wide-area  oscillations, such as between Ukraine and Spain,  and between  Sweden and Spain \cite{danilo}. The problem of forced oscillations has also been studied \cite{Slava}. 

As these problems continue to emerge, it has become very difficult to identify their root causes,  and to design control needed to suppress them.  Most of the efforts have gone into transient stability simulations  and  small signal stability analyses, in particular as the slow power plants get replaced by smaller-scale Inverter Based  Resources (IBRs)  \cite{danilo}. A particular new challenge perceived by the industry  is a lack of detailed models of diverse components.  Also, several renown researchers have proposed   in the past modeling and analyses approaches, notably use of normal forms {\cite{nnaemeka}} and notions of dissipating energy flows \cite{Slava2}. { While these methods help with locating the source of inter-area oscillations, they do not model the inter-area interactions dynamically.}

The primary contribution of this paper is to meet the need to have a physically-intuitive interpretation of dynamic interactions, in terms of energy conversion, power and rate of  power  exchanges, between different energy conversion components and their relations to the existence of inter-area oscillations within an interconnected electric energy system.  We do this  by observing that each stand-alone module $i$ (component, sub-system) can be characterized in terms of its own internal states and  the shared  \cite{willems} interaction variable $intVar^i$ which is a function of its own internal states only, as introduced  in our  earlier work \cite{naps1,naps2}.
{We then provide a notion of dynamic interaction between two areas by introducing, for the first time, an inter-area variable as the difference of the interaction variables of the respective areas.}

In Section \ref{sec:2} we  review a previously introduced  aggregate state space model representing dynamics of energy and power within  a  multi-area  interconnected electric power system; we review a previously-introduced   structural  definition of a technology-agnostic  interaction variable associated with each   stand-alone module $i$;  and,  define electro-mechanical  inter-area oscillation as an  interplay between interaction variables of neighboring modules $j \in C_i$. We stress that these concepts are  general and applicable to systems comprising both conventional  generation and renewable resources.

Next, in Section \ref{sec:3}, we illustrate a particular case of linearized power-frequency energy dynamics comprising conventional power plants. This section is a short summary of the earlier work \cite{naps1}  which laid the foundations for claiming the existence  of interaction variables in linearized models of electric power grids.  An important aspect of this modeling of power  dynamics is that the load power deviations from forecast can be explicitly modeled. 

Section \ref{sec:4}  concerns  eigenvalue  analyses of states contributing to inter-area oscillations. To start with, the existence of an interaction variable associated with a stand-alone module is a direct result of its own power conservation. 
Importantly, the structural existence of an eigenmode in an interconnected system, not present in two disconnected subsystems, is now shown to be a direct consequence of power conservation, both mathematically and through simulations.  Based on these properties,  in sub-section \ref{sec:5} we propose a method for quickly identifying the modules whose states contribute the most to the inter-area oscillations.  This method  could be  potentially of major use in real-world large-scale systems for which understanding causes of inter-area oscillations becomes a major problem \cite{danilo}.

In Section \ref{sec:6} we  describe numerical simulations to illustrate the sensitivities of inter-area oscillations with respect to the strength of tie-line flow  connections and inertia of generators. For the first  time  the effects of grid design and  energy conversion processes on physically-intuitive  properties  such as frequency of inter-area oscillations can be interpreted without requiring full-blown  time domain simulations.  

{In Section \ref{sec:7} we add an intermittent source and simulate an example to show its effect on the inter-area oscillations depending on the frequency of intermittent disturbances.}
We close in Section \ref{sec:8} by stressing the structural nature of inter-area oscillations  and potential for further generalization of these properties to multi-energy systems. 

\section{ Structural  property of complex multi-area electric power systems}
\label{sec:2}
\begin{figure} [h]
    \centering
    \includegraphics[width=0.65\columnwidth]{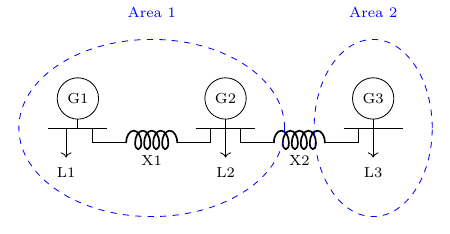}
    \caption{An interconnected 3-bus 2-area system}
    \label{fig:3bus2area}
\end{figure}

In electric  power systems, such as the one shown in Figure \ref{fig:3bus2area}, typically single port modules, generators and loads, are interconnected via transmission lines whose real power flows  are nonlinearly dependent on the nodal phase angle differences. 
All components, except perfect resistors,  have  state variables and  time constants defining their  natural   and/or forced response  given initial conditions. In today's AC electric power systems these state variables are  frequency $\omega(t)$, voltage $V(t)$, real power $P(t)$ and reactive power $Q(t)$. 
When writing the dynamic model in terms of these variables,  one quickly loses physical insights. It therefore becomes quite hard to understand root causes of dynamic problems of interest. To overcome this complexity,  we take a modular approach to modeling complex interconnected electric power systems.
We introduce a transformed state space which is derived  by combining the dynamics of conventional internal state variables $\tilde x_{old,i} (t)$ in each module $i$  (except the nodal phase angle $\delta_i$) and its generated power $P_i(t)$.  The structural form of this transformed state space is derived by introducing a new, transformed state variable comprising  a technology-dependent   $\tilde x_{old,i}(t)$ and the power $P_i(t)  = \frac {dE_i(t)} {dt}$ 
 \begin{align}
 x_{new,i}(t) = \begin{bmatrix}
     \tilde x_{old,i}(t) & P_i (t)
 \end{bmatrix}^\top    
 \label{eqn:xnew}
 \end{align}  

\subsection{General structure of single port dynamic components}

We first derive a general structure of single port components by recognizing that each one  can be thought of as an energy conversion processing module which has certain structural properties resulting from fundamental conservation of energy. 
\begin{equation}
    \dot E_i (t)  = - \frac{E_i(t)}{\tau_i} - P_i(t) + P_{in, i}(t), \;\; E_i(0) = E_{i0}
    \label{eqn:enconv}
\end{equation}
Here $E_i(t)$, and $P_i(t)$ are stored energy in module $i$ and real power generated by module $i$ at the port/node.
$P_{in, i}(t)$ is the exogenous power input into module $i$. The term $\frac{E_i(t)}{\tau_i}$ stands for Joule losses internal to module $i$, and $\tau_i$ is the time constant representing rate at which energy stored in  the module dissipates, and can be interpreted as  the ratio of stored and dissipated energy \cite{arcrupa}. 

We also  recall that the rate of change of energy conversion within  each module  $\dot E_i(t)$  is technology-specific and can be expressed in terms of  derivatives of its own local state variables $\dot{\tilde{x}}_{old,i}(t)$.  It then follows from the basic conservation law  in \eqref{eqn:powbal} when combined with energy conversion \eqref{eqn:enconv}  that  the  dynamics of the local variables $\tilde x_{old,i}(t)$  can be expressed  structurally in terms of its own states and its rate of change of energy, namely power.
\begin{equation}
      \dot{\tilde{x}}_{old,i}(t) = \tilde f_i (\tilde x_{old,i}(t), P_i (t), u_i(t), m_i(t))
      \label{eqn:xolddyn}
\end{equation}
where $u_i(t)$ and $m_i(t)$ are the  local control and  exogenous inputs,  respectively.
The key observation when interconnecting modules of diverse energy conversion components is that power and rate of change of power are Tellegen's quantities, as a direct consequence of power conservation \cite{penfield}. This means that power output from module $i$ must equal power injected into modules $j \in C_i$, where $C_i$ is the set of $i$'s neighbouring modules. For the example of electric power grid, power produced by the  generator  equals the sum of power taken out by the components $j$ directly connected to it, namely  transmission lines.  
\begin{equation} P_i(t) = \sum_{j \in C_i} P_{ij}(t)
\label{eqn:powbal}
\end{equation}
where $P_{ij}(t)$ is the real power flow from module $i$ to $j$.
For purposes of capturing rates of interactions between modules, we observe that rates of change of power $\dot P_i(t)$  must balance with the rates of change of power, say $\dot P_{ij}(t)$, taken out  by the other components, $j \in C_i$, 
\begin{equation}
\dot P_i(t) = \sum_{j \in C_i} \dot P_{ij}(t)
\label{dotpowbal}
\end{equation}

The dynamics of $x_{new,i}$ takes on the form, 
\begin{align}\label{eqn:newstates}
    \dot{x}_{new,i}(t)  & = \begin{bmatrix}
        \dot{\tilde{x}}_{old,i}(t) \\ \dot P_i(t)
    \end{bmatrix} =
    \begin{bmatrix}
        \tilde f_i (\tilde x_{old,i}(t), P_i (t), u_i(t), m_i(t) \\h_i(\tilde x_{old,i}(t), \dot P_j, j \in C_i)
    \end{bmatrix}
\end{align}


\subsection{Existence of interaction variable at component level}
{ Consider a component's interaction variable as the power generated by the component $P_i$ \cite{naps1, naps2}.} 
By re-writing \eqref{eqn:enconv} as
\begin{equation}
    P_i (t)= P_{in,i} (t) -\dot E_i(t) -\frac {E_i(t)} {\tau_i} 
\label{eqn:intvarcomp}
\end{equation}
it follows that a disconnected component has $P_i(t)=0$ and that it is a function only of its internal inputs and state variables.

\subsection{Simplified model of two-port components}

Similarly, the basic conservation of power for two-port transmission lines interconnecting its sending end $i$ and receiving end $j$ is   
\begin{equation}
   \frac {dE_{ij}(t)}{dt} = P_i(t) +P_j(t) -\frac {E_{ij}}{\tau_{ij}}
   \label{eqn:trline}
\end{equation}
where $\frac {E_{ij}} {\tau _{ij}}$ represents Joule losses in the transmission line, and, assuming negligible resistance it is approximately zero. When studying slow electro-mechanical oscillations  $ \frac {dE_{ij}(t)}{dt} \approx 0$ because the time constants of its LC parameters are  near zero  relative to the time  constants of electrical machines.  Consequently, the line is modeled as having inductance parameter $X_{ij}$. It follows from \eqref{eqn:trline} that 
\begin{equation}
    P_i(t) +P_j(t) =0 
   \label{eqn:trline_instant} 
\end{equation}

\subsection{Interaction variable of stand-alone area $i$}
To review a general structural notion of an interaction variable $intVar^1$ of Area  1 proposed in \cite{naps1,naps2},  
we re-state constraint  given in   \eqref{eqn:powbal}
 for each network node within the  area by stating a conservation of rate of change of power for the two modules (generator-load pairs are taken as one module) in   $Area\; 1$ 
 \begin{equation}
      P_{i} = \sum_{j \in C_i}  P_{ij}(t)   
     \label{eqn:dotPGi}
 \end{equation} 
For all $i \in Area \; 1$ in  the network shown in Figure \ref{fig:3bus2area} power generated by the two modules at the two nodes are $P_1^1(t) = P_{G1}^1(t)-P_{L1}(t)$, and $P_2^1(t) = P_{G2}^1(t)-P_{L2}(t)$, respectively. 
Further, when applying constraint given in \eqref{eqn:powbal} we get,  
\begin{equation}
\dot z^1(t) = {intVar}^1(t) =   P_1^1(t) + P_2^1 (t) = F_{1-2}
\label{def:intvar}
\end{equation}
where $F_{1-2}$ is the power flow from Area 1 to Area 2 through the tie-line. 
We can infer that, for a disconnected area, 
$F_{1-2}(t)=constant$
and $\dot F_{1-2}(t)=0$, i.e.
net power out of disconnected area  must remain  constant and its derivative is zero. This linear combination of net power generated by all modules within a given Area 1 is defined as the $intVar^1(t)$ of Area 1 with the rest of the system \cite{naps1,naps2}.

\subsection{Inter-area dynamics: Definition in terms of interaction variables of stand alone areas}
We propose, given the above fundamental relations for energy conversion, and the  notion of stand-alone of interaction variable $intVar^I (t)$, 
 for each area $I$ that when  interconnecting the two areas shown in Figure \ref{fig:3bus2area} via transmission line, whose parameter is $X_{2}$, general Tellegen's Theorem in \eqref{eqn:powbal} implies: 
\begin{itemize}
    \item Incident power  into transmission line from Area $1$ must be the same as  the power  generated by the Area $1$, given in \eqref{eqn:intvarcomp}, $I=1$.
    \item  Reflected power  into transmission line from Area $2$ is given by the same \eqref{eqn:intvarcomp},  $I= 2$
   \item {\textbf{Definition:   Inter-area variable I-II}} is the difference between incident and reflected powers into the tie line interconnecting them.
\end{itemize}
 It can be seen from  \eqref{eqn:intvarcomp} that the incident power is mainly contributed  by some internal power injected into Area 1 reduced by the Joule losses inside the area and the rate for change of energy consumed  by the area.  These definitions of incident and reflected power are degenerate since the travelling waves of tie-line are not modeled when capturing slow electro-mechanical inter-area dynamics which is mainly created by the difference in powers injected from the areas.

\section{Linearized modeling  of  power dynamics}
\label{sec:3}
In this section we suggest that the linearized modeling of power dynamics introduced earlier in \cite{naps1} follows from the more general derivations  in Section \ref{sec:2} which do not require linearization and are technology-agnostic. For completeness, we  briefly summarize this early derivation. To do so, we  
consider the same two-area interconnected system  shown in Figure \ref{fig:3bus2area}, which comprises a mix of conventional generators and loads interconnected via  lossless transmission lines. 
%

\subsection{Dynamics of a single generator}
The linearised model of a single governor-turbine-generator (G-T-G) set includes the rotor swing equation, 
the turbine dynamics and the local governor feedback proportional control.  
The standard state space form of a  governor-turbine-generator (G-T-G) \eqref{eqn:statespaceLC} has the generator state $\omega_G$, the turbine state $P_t$ and the governor state $a$, all being the  deviations from their nominal value.
\begin{align}
    \dot{x}_{LC} &= \begin{bmatrix} \dot{\omega}_G \\ \dot{P}_t \\ \dot{a}
    \end{bmatrix} = \begin{bmatrix}
        \frac{-D}{M} & \frac{1}{M} & 0 \vspace{0.15cm}\\ 
        0 & \frac{-1}{T_t} & \frac{K_t}{T_t} \vspace{0.15cm}\\
        \frac{-1}{T_g} & 0 & \frac{-1}{rT_g}
    \end{bmatrix}  \begin{bmatrix} {\omega}_G \\ {P}_t \\ {a}
    \end{bmatrix} + \begin{bmatrix}
        \frac{-1}{M} \\ 0 \\ 0
    \end{bmatrix} P_G,  \nonumber\\
    & = A_{LC} x_{LC} + c_M P_G, \label{eqn:statespaceLC}
\end{align}
where $M, D, T_t, T_g$ are moment of inertia of the combined G-T-G set, its damping coefficient, and the time constants of the turbine and governor, respectively. 
$\omega_G, P_G, P_t, a$ are the rotor frequency of the generator, its real power output, power produced by the turbine, and the valve opening, respectively. 
$K_T$ and $r$ are the local control gains in the turbine and governor respectively. This model is effectively a particular case of  the general nonlinear dynamics of a generator \eqref{eqn:newstates} in transformed state space.


\subsection{Network constraints}
Real power flow into a node is a function of the nodal voltage magnitudes and phase angles of the network, $P^N = P^N(\delta, V)$. 
Linearizing power flow into a generator bus and a load bus under the decoupling assumption ($\partial P^N / \partial V = 0$) gives us,
\begin{align}
    P_G + F_G &= J_{GG} \delta_G + J_{GL} \delta_L \label{eqn:vectorgen}\\
    F_L - P_L &= J_{LG} \delta_G + J_{LL} \delta_L \label{eqn:vectorload}\\
    \nonumber
\end{align}
where the network has $n$ generators and $m$ loads. 
$\delta_G$ and $\delta_L$ correspond to the vector of phase angles, {and $F_G$ and $F_L$ correspond to the real power flow from neighbouring modules} at the generator and load bus respectively.
The network constraints given  in \eqref{eqn:vectorgen} and \eqref{eqn:vectorload}  are well-known linearized power flow equations, and represent a particular case of a more general power balance given as \eqref{eqn:powbal} earlier. 
The main step in deriving a transformed state space requires applying  general constraints on rate of change of power shown in \eqref{dotpowbal} introduced earlier, implying that differentiation of power balance equations is a valid step. In today's power systems problems $P_L$ is predicted and dynamical system response to its deviations $\dot P_L$  is of direct interest.  This is obtained by expressing $\omega_L$ as a disturbance caused by the load power deviations $\dot P_L$ and  replacing it in the  network constraints imposed on the rate of change of  power generation $\dot P_G$
In short, define
\begin{align}
    K_P = J_{GG} - J_{GL}J_{LL}^{-1}J_{LG}, \;\; 
    D_P = J_{GG}J_{LL}^{-1}
\end{align}
Differentiate and rearrange \eqref{eqn:vectorgen} and \eqref{eqn:vectorload}, to obtain  
\begin{align}
    \dot{P}_G &= K_P \omega_G + \dot{F_e} - D_P \dot{P}_L
\label{eqn:genmodel}
\end{align}
Here $F_e = J_{GG}J_{LL}^{-1}{F}_L - F_G$ represents the effective flow from neighbouring areas as seen by each generator.
Equations \eqref{eqn:statespaceLC} represent linearized model of G-T-G components in the entire system   and  when subject to Tellegen's constraints   on their rate of change power in the interconnected system,  given as \eqref{eqn:genmodel}, jointly result in a particular case of transformed state space model introduced earlier in \eqref{eqn:newstates}.

\subsection{Interconnected system dynamics in transformed state space}
It was shown in \cite{naps1,naps2} that when using the  vector and diagonal matrix  notation for a general power system,   
and combining the local generator dynamics and the network constraints, one obtains without loss of generality  transformed state-space model of the system \cite{naps1,naps2}
\begin{align}
    \begin{bmatrix} 
        \dot{x}_{LC} \\ \dot{P}_G
    \end{bmatrix} = 
    \begin{bmatrix}
        A_{LC} & C_M \\ K_P E & 0
    \end{bmatrix}
    \begin{bmatrix}
        {x}_{LC} \\ {P}_G
    \end{bmatrix} + 
    \begin{bmatrix}
        0 \\ \dot{F}_e
    \end{bmatrix} +
    \begin{bmatrix}
        0 \\ -D_P \dot{P}_L
    \end{bmatrix} \label{eqn:transformedss}
\end{align}
The system matrix of the augmented state space takes on the form 
\begin{align}
    A = \begin{bmatrix}
        A_{LC} & C_M \\ K_P E & 0
    \end{bmatrix}
    \label{eqn:matrix}
\end{align}
This model is basically a linearized model of a more general nonlinear model derived earlier in \eqref{eqn:transformedss}. 
\section{Eigen-mode Analysis of Inter-area Dynamics}
\label{sec:4}

In  this section, we make use of the transformed state space model shown above  and its system matrix given in  \eqref{eqn:matrix} to  carry our eigen-mode analysis. In particular, given structural properties of this model and its system matrix, we wish to assess relations between the eigen-modes in disconnected control areas as well as in the  interconnected system.  We expect that each disconnected area itself has a zero eigen-mode reflecting structural constraint on its Tellegen's quantities, power and/or rate of change of power. These modes are naturally contributed by the states internal to the areas. 
We also anticipated that an additional eigen-mode appears when the areas are interconnected and that the state contributing to this eigen-mode  are from both areas. The inter-area oscillations (IAO) are then contributed by these states. 
Using participation factor analysis  we identify the state variables of the components that contribute the most to the IAO eigen-mode. Most importantly, we illustrate how the transformed state space makes it convenient to identify causes of oscillations.

\subsection{Methodology to identify the interconnection eigen-mode}
\label{sec:5}
To understand which states contribute the most to the inter-area oscillations, we first need to identify which eigen-modes of the system arise due to the interconnection of the areas. 
A general method to identify the interconnection eigen-mode is explained below, followed by a simulated example. 

For our analysis, consider an undamped single isolated system ($F_e = 0$) with constant power load ($\dot{P}_L = 0$).
Its state space formulation reduces to,
\begin{align}
    \dot{x} = A x
\end{align}
We simulate two cases, 
\begin{itemize}
    \item \textbf{Connected Isolated System (CIS):} The original system of interest, like in Figure \ref{fig:3bus2area}
    \item \textbf{Disconnected Isolated System (DIS):} The same system in Figure \ref{fig:3bus2area} but with a disconnected inter-area tie line
\end{itemize}
The structure of the eigen-values of $A$ \cite{naps1,naps2} for DIS is,
\begin{align}
    \lambda(A_{DIS}) = \begin{bmatrix}
        0 \\ \vdots \\ 0 \\ \vdots\\ \pm \Omega_{1} j
    \end{bmatrix}
\end{align}
where $\pm \Omega_{1} j$ corresponds to the local eigen-values of the two areas.
The zero eigen-values are due the conservation of power/rate of change of power in each of the areas.
The structure of the eigen-values of $A$ \cite{naps1,naps2} for CIS is,
\begin{align}
    \lambda(A_{CIS}) = \begin{bmatrix}
        0 \\ \vdots \\ \pm \omega j \\ \vdots \\ \pm \Omega_{2} j
    \end{bmatrix}
\end{align}
where, like in DIS, $\pm \Omega_{2}j$ corresponds to the local eigen-values of the two disconnected areas. 
We have an additional non-zero eigen-value $ \pm \omega j$ which arises due to the interconnection of the two areas and is responsible for the IAO. 
To identify the eigen-value that corresponds to the interconnection, one can compare $\lambda_{DIS}$ and $\lambda_{CIS}$ and find the additional non-zero interconnection eigen-value $\pm \omega j$.  In most cases, the interconnection eigen-mode will be of a lower frequency than the local eigen-modes.
A simple participation factor analysis can then be done to find the state variables contributing to $\pm \omega j$.

\section{Sensitivity of inter-area oscillations with respect to system parameters}
\label{sec:6}
We simulate an example to better understand the effect of weak inter-area tie lines and low inertia generators on inter-area oscillations (IAO). 
Consider the three-bus two-area system in Figure \ref{fig:3bus2area}. 
The system is undamped, isolated and is subjected to constant power loads $L_1, L_2,$ and $L_3$. 
The lines are taken as purely inductive with reactances $X_1$ and $X_2$.
A similar analysis can be done for a damped system, but the idea remains the same.

\subsection{Effect of tie-line strength} \label{subsec:tieline}
We demonstrate the effect of the strength of the inter-area tie-line on the interconnection eigen-mode and IAO by considering three cases.
\begin{enumerate}
    \item \underline{Case 1:} 
    We keep the reactance of tie-line relatively small and comparable to the local line strengths. 
    $$X_2 = X_1$$
    \item \underline{Case 2:} 
    We keep the reactance of tie-line relatively bigger than the local line strengths. 
    $$X_2 = 10 X_1$$

\item \underline{Case 3:} 
    This is the DIS case, in Figure \ref{fig:3bus2area}, where the tie-line is very weak and almost non-existent.
    $$X_2 >> X_1$$
\end{enumerate}

In our eigen-mode and participation factor analysis for case 1, we identify the following non-zero eigenvalues and their primary contributing state variables.
\begin{align*}
    \Omega^1 &= \pm 3.7499 j \longrightarrow \omega_{G}^2, P_G^2\\
    \omega^1 &= \pm 2.1650 j \longrightarrow \omega_{G}^1, P_G^1, \omega_{G}^3, P_G^3
\end{align*}
Similarly for case 2,
\begin{align*}
    \Omega^2 &= \pm 3.1028 j \longrightarrow \omega_{G}^1, P_G^1, \omega_{G}^2, P_G^2\\
    \omega^2 &= \pm 0.8274 j \longrightarrow \omega_{G}^1, P_G^1, \omega_{G}^3, P_G^3
\end{align*}
And for case 3,
\begin{align*}
    \Omega^3 &= \pm 3.0618 j \longrightarrow \omega_{G}^1, P_G^1, \omega_{G}^2, P_G^2
\end{align*}
It is straightforward to identify the interconnection eigen-mode for case 1 and case 2 as $\omega^1$ and $\omega^2$ respectively. 
The states of generator $G_1$ and $G_3$ contribute to the interconnection eigen-mode, the mode that disappears in the DIS case 3. 
We also note that the local eigen-modes, $\Omega^1, \Omega^2, \Omega^3$, are primarily dependent on the states of generators in their respective area. 

\begin{figure}
    \centering
    \includegraphics[width=\columnwidth]{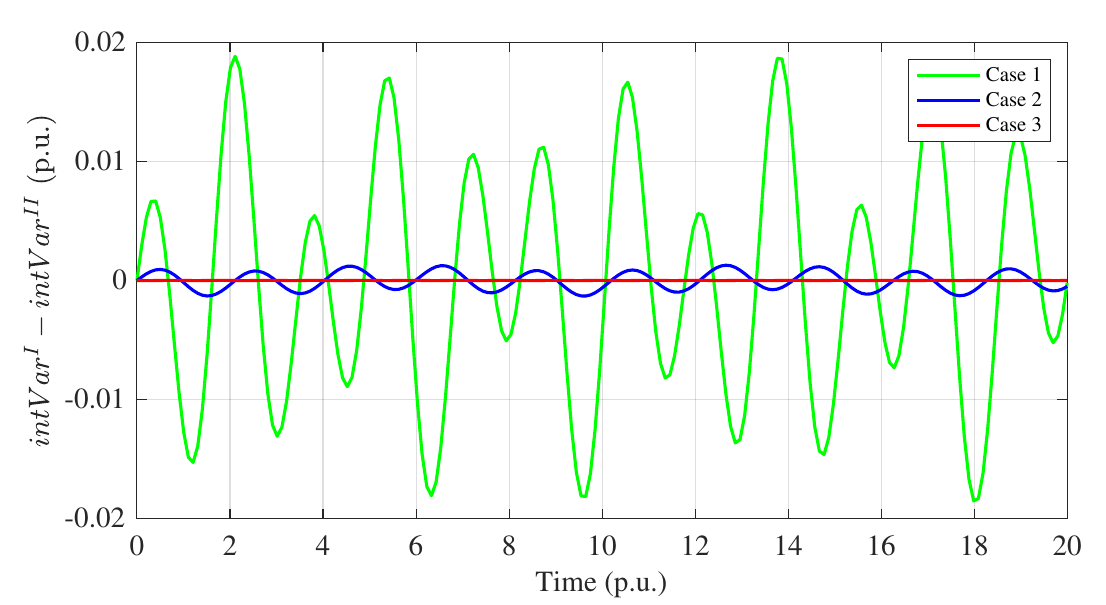}
    \caption{Inter-area oscillation as a difference of  the rate of change interaction variables in  the two areas for varying tie-line strength}
    \label{fig:z}
\end{figure}

As defined in Section \ref{sec:2}, inter-area oscillations are expressed as the difference between the rate of change of interaction variables of two  disconnected areas.
We plot this for the three cases in Figure \ref{fig:z}. 
The weaker the line gets, the slower is the oscillation of the power flowing through it, with the oscillation going to 0 for DIS. 
This can also be seen in the eigenvalues obtained for each case.


\subsection{Effect of inertia of generator}\label{subsec:inertia}

We now demonstrate the effect of generator inertia on the interconnection eigen-mode and IAO. 
Again, consider the isolated system in Figure \ref{fig:3bus2area} and take two cases.
\begin{enumerate}
    \item \underline{Case 1:} The inertia of all generators is equal
    $$M_{G1} = M_{G2} = M_{G3}$$
    \item \underline{Case 2:} The inertia of generator in area 2 is taken to be relatively bigger than the other generators
    $$M_{G1} = M_{G2} < M_{G3}$$
\end{enumerate}
We take $X_1 = X_2$ for this particular experiment. The parameters chosen for our simulation are given in the appendix. 
On doing eigen-mode analysis and participation factor analysis for case 1 we get,
\begin{align*}
    \Omega^{1} &= \pm 3.7499j \longrightarrow \omega^2_G, P_G^2\\
    \omega^1 &= \pm 2.1650j \longrightarrow \omega^1_G, P_G^1, \omega^3_G, P_G^3
\end{align*}
Similarly for case 2,
\begin{align*}
    \Omega^{2} &= \pm 3.5221j \longrightarrow \omega^1_G, P_G^1, \omega^2_G, P_G^2\\
    \omega^2 &= \pm 1.4578j \longrightarrow \omega^1_G, P_G^1, \omega^2_G, P_G^2, \omega^3_G, P_G^3
\end{align*}

\begin{figure}
    \centering
    \includegraphics[width=\columnwidth]{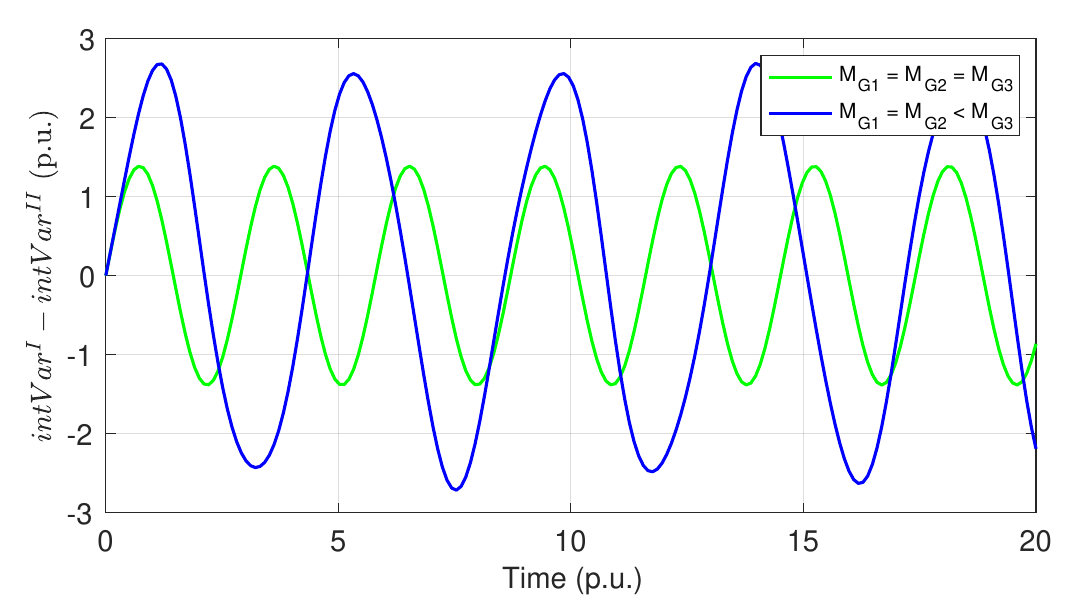}
    \caption{Inter-area oscillation as a difference of rate of change of  interaction variables from two areas for varying generator inertia}
    \label{fig:zinertia}
\end{figure}

When $G_3$ has a large inertia, the interconnection eigen-mode ($\omega^2$) gets slower and $G_3$ states start contributing more to it.
As is expected, generators with large inertia have a slower response and can be seen in the Figure \ref{fig:zinertia} which plots the inter-area oscillation for the two cases.

\section{Impact of Intermittent Resources on IAO}
\label{sec:7}

\begin{figure}
    \centering
    \includegraphics[width=\columnwidth]{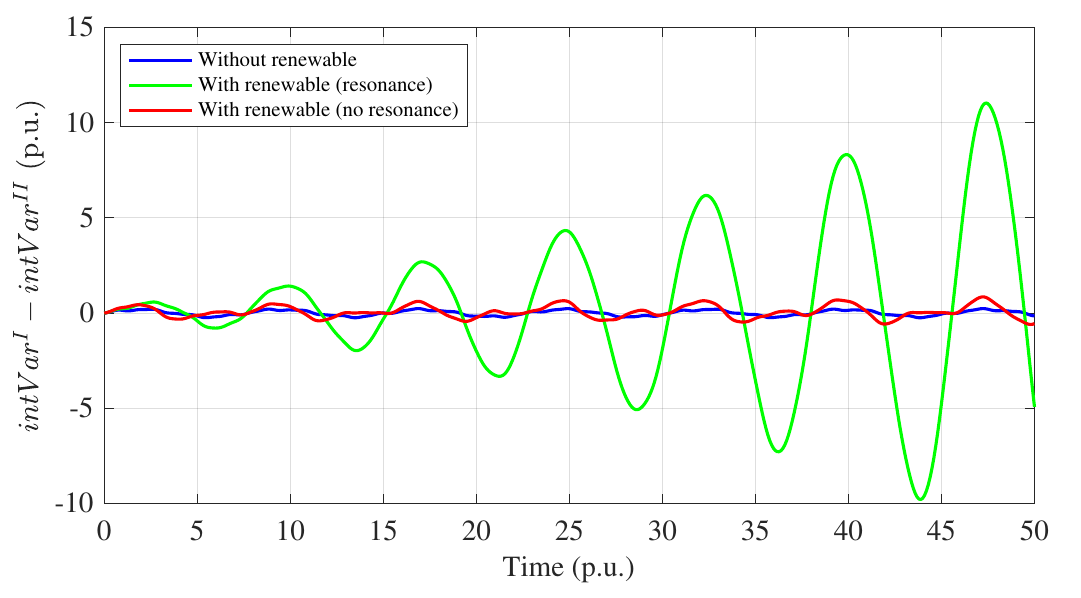}
    \caption{Inter-area oscillation expressed as a difference of rate of change  interaction variable of the two areas with and without a varying renewable source}
    \label{fig:renewableplot}
\end{figure}

The transformed state space has another advantage; it allows us to express load disturbances in the form of $\dot{P}_L$ \eqref{eqn:transformedss}. 
Specifically, renewable sources are intermittent and can be characterised as a varying negative load. 
Consider the network in Figure \ref{fig:3bus2area}, with and without a renewable source in the place of $L_1$. 
When the damping is insufficient and the renewable power output deviations include the interconnection mode ($\dot{P}_L = A_\circ\sin(\omega t)$, $\omega$ is the interconnection eigenvalue), it resonates with the IAO and makes it unstable as can be seen in Figure \ref{fig:renewableplot}.

\section{Concluding remarks}
\label{sec:8}

The notion of  $intVar$ used in this paper has two unique properties: when the area is disconnected $intVar$ remains constant, and, its rate of power exchange with other areas is zero.
Given these properties, we claim  that the energy dynamics interpretation of interactions within a multi-area system  is structural, {i.e., they depend only on the basic conservation laws and Tellegen's theorem and are agnostic to technology of the internal energy conversion dynamics of modules. This means that the inter-area dynamics only depend on the range of interface specifications, an essential property for distributed modeling, sensing, computing and control.}
An example of  aggregate mathematical model is shown on a small electric power system   comprising two areas to illustrate these structural  properties. {Future work include, incorporating real-reactive power coupling and losses in the modeling, analysing practical examples} and studying the relationship between feedback linearizing control and the structural inter-area property for suppressing slow inter-area oscillations between New York and New England areas \cite{king} and for eliminating sub-synchronous resonance \cite{allen}.  
\begin{appendix}
    For Section \ref{subsec:tieline}, all generators are taken to be identical, i.e., their $A_{LC}$ matrix is the same. 
    The generator inertias are $M = 3.2 \; \text{p.u.}$ and line reactance $X_1 = 6.67 \cdot 10^{-2}\; \text{p.u.}$.
    For Section \ref{subsec:inertia}, $X_1 = X_2 = 6.67 \cdot 10^{-2}\; \text{p.u.}$. 
    For case 1, generator inertias are $M_{G1} = M_{G2}= M_{G3}= 3.2 \; \text{p.u.}$ and for case 2, $M_{G3} = 32 \;\text{p.u.}$. 
    All simulations show deviations of the quantities around equilibrium.
    As both the examples consider an undamped system, $D = 0$ and the time constants $T_t, T_g$ are very large numbers. {}
\end{appendix}

\end{document}